\documentclass[12pt,a4paper]{ijicic}
\usepackage[dvipdfm]{graphicx}
\usepackage{amsmath,amssymb}
\usepackage{cite}

\makeatletter
\def\ave#1{\langle#1\rangle}
\def\lfig#1{\label{fig:#1}}\def\rfig#1{Fig.\ref{fig:#1}}
 \def\Rfig#1{Figure \ref{fig:#1}}
\def\ltab#1{\label{tab:#1}}\def\rtab#1{Table \ref{tab:#1}}
 \def\Rtab#1{Table \ref{tab:#1}}
\def\lsec#1{\label{sec:#1}}\def\rsec#1{\ref{sec:#1}}
\def\leqn#1{\label{eqn:#1}}\def\reqn#1{(\ref{eqn:#1})}
\def\bi#1{\ensuremath{\boldsymbol{{#1}}}}

\def\@pd[#1]#2#3{{\def\aaa{#1}\ifx\aaa\@empty%
 \frac{\partial #2}{\partial {#3}}%
 \else\frac{\partial^{#1} #2}{\partial {#3}^{#1}}\fi}}
\def\@od[#1]#2#3{{\def\aaa{#1}\ifx\aaa\@empty%
 \frac{d #2}{d #3}%
 \else\frac{d^{#1} #2}{d {#3}^{#1}}\fi}}
\def\pd{\@ifnextchar[{\@pd}{\@pd[]}}
\def\od{\@ifnextchar[{\@od}{\@od[]}}
\def\Dx{\,\Delta x}
\def\Dq{\,\Delta q}
\def\Dt{\,\Delta t}
\def\dDx{\,\Delta \dot x}
\def\ddDx{\,\Delta \ddot x}
\def\dDq{\,\Delta \dot q}
\def\ddDq{\,\Delta \ddot q}
\def\STCC#1#2{R(#1,#2;\tau)}
\makeatother

\title{Effects of Mechanical Coupling on the Dynamics\\
of Balancing Tasks}

\def\currentvolume{7}
\def\currentissue{3}
\def\currentyear{2011}
\def\currentmonth{March}

\begin{document}
\maketitle

\centerline{\scshape  Katsutoshi Yoshida}
 \medskip
{\footnotesize
\centerline{Department of Mechanical and Intelligent Engineering}
\centerline{Utsunomiya University} 
\centerline{7-1-2 Yoto, Utsunomiya-shi, Tochigi 321-8585, Japan}
\centerline{yoshidak@cc.utsunomiya-u.ac.jp} }
\medskip

\centerline{\scshape  Atsushi Higeta}
 \medskip
{\footnotesize
\centerline{Department of Innovation Systems Engineering}
\centerline{Utsunomiya University} 
\centerline{higeta@katzlab.jp} }
\medskip

\centerline{\scshape  Shinichi Watanabe}
 \medskip
{\footnotesize
\centerline{Innovation Center for Research and Engineering Education}
\centerline{Utsunomiya University} 
\centerline{swtnb@cc.utsunomiya-u.ac.jp} }
\medskip

\maketitle

\begin{abstract}
 Coupled human balancing tasks are investigated based on both
 pseudo-neural controllers characterized by time-delayed feedback with
 random gain and natural human balancing tasks.  It is shown numerically
 that, compared to single balancing tasks, balancing tasks coupled
 by mechanical structures exhibit enhanced stability against balancing errors in
 terms of both amplitude and velocity and also improve the tracking
 ability of the controllers. We then perform an experiment in which 
 numerical pseudo-neural controllers are replaced with natural human
 balancing tasks carried out using computer mice.  The results reveal that
 the coupling structure generates asymmetric tracking abilities in
 subjects whose tracking abilities are nearly symmetric in their
 single balancing tasks.
\end{abstract}

\section{Introduction}

Competitive and cooperative dynamics can arise when multiple agents
(autonomous entities) share common resources and environments.
Extensive research has been conducted on such mutual interactions.
Research in this area can be broadly classified into two categories: the
field of mathematical ecology \cite{hofbauer01}, which finds group
behavior models having low degrees of freedom, and the field of
multi-robot systems \cite{DARS.en}, which develops individual agents
generating group behavior.
 In these conventional approaches, couplings between agents are
 constructed using shared resources or information.

 On the other hand, it is also reasonable to consider another type of
 coupling, constructed of mechanical structures, for instance,
 connection rods placed between agents. Such a mechanical type of
 coupling appears to play an important role in clarifying the dynamics
 of agents, including humans and robots, which are in physical contact
 with one another. This type of knowledge will be applicable to
 developing robots working together or robots supporting human
 activities, based on physical contact.  However, it appears that little
 is known about the type of dynamics that is produced by mechanical
 coupling.
 Therefore, as a preliminary step toward solving this problem,
we have developed the coupled inverted pendula (CIP) model with four
DOF \cite{JSDD08}, in which nonlinear behavior similar to interspecific
competition in an ecosystem \cite{hofbauer01} can be directly generated
by individual mechanical structures of agents.  The CIP model proposed
by the authors consists of a pair of independently PD--controlled
inverted pendula, the tips of which are connected through a rigid rod.

 The primary approach used in the present study is to replace the PD controllers of the
 CIP model with human balancing tasks. It follows that the two subjects
 perform stick balancing tasks under the situation in which the tips of the 
 sticks are mechanically connected to each other.
 This approach will allow us to determine how humans work together dynamically
 through physical contacts and will consequently provide helpful information 
 for developing robots to support human activities.

 From an engineering point of view, stick balancing is basically
 achieved by the tracking control technique (See
 References~\cite{Ming2009,Chien2009} for typical examples in
 engineering problems).  However, human tracking control is known to
 exhibit further characteristics beyond engineering tracking
 control.
 For example, the human tracking control is known to have
 a relatively large time-delay caused by neural latency and parametric
 random fluctuations typical of neural control\cite{Cabrera02}.
In the field of statistical mechanics, it has been reported that random
fluctuations arising in single human stick balancing tasks can be
accurately modeled as an inverted pendulum with a time-delayed and
randomly modulated feedback controller \cite{Cabrera02}.  Since such a
controller does not simulate the structure of neural
networks\cite{campbell04,campbell06}, it should be referred to as a
pseudo-neural control model of human balancing tasks.  The most
remarkable effect of the pseudo-neural controller is that, near
stability boundaries, parametric noise can allow the controller to
produce corrective movements on time scales shorter than the delay time
of the controller under certain suitable conditions, exhibiting the
scaling laws typical of the self-similarity dynamics of on-off
intermittency \cite{Shankar96}. This effect has also been experimentally
confirmed in regard to physical human tasks, such as stick balancing
\cite{Cabrera02,Cabrera04} and visuomotor tracking on a computer screen
\cite{Bormann04}.
 In these studies on the human balancing task, however, the case for two
 coupled humans has never been considered. Although common features of
 human control would remain even in the coupled case, there must be
 specific effects related to interactions between the two humans. This
 problem remains to be addressed.

 In the present paper, we consider the open problem of what kind of
 dynamics may arise when mechanical coupling is imposed on the balancing
 tasks by two humans whose dynamics can be characterized by the
 time-delayed and randomly modulated feedback controller.
 To achieve this,
we propose a new model consisting of time-delayed and randomly modulated
feedback controllers that are coupled by a mechanical structure
equivalent to the proposed CIP model. For simplicity, we first derive a
linearized reduced--order version of the CIP model, after which we
replace the PD controllers with pseudo-neural controllers, as developed
in the literature \cite{Cabrera02,Cabrera04,Bormann04}. Using this
model, we demonstrate the effects of coupling, namely, an improvement in stability
and sensitivity of the corrective motions in the balancing tasks.  We
also perform an experiment in which we replace the numerical controllers
with human visuomotor tracking tasks performed by subjects using
computer mice. It is shown experimentally that the coupling structure
between two subjects induces asymmetric sensitivity in the corrective
motions of the subjects.

\section{Analytical Model}

\subsection{Single balancing tasks}

Human stick balancing tasks and their scaling properties have been
accurately modeled as an inverted pendulum with a time-delayed feedback
of random gain of the following form \cite{Cabrera02}:
\begin{equation}
 \ddot \theta + \gamma\dot \theta - \alpha\sin\theta 
  + \beta\,R(t)\theta(t-\tau) = 0
  \leqn{cabrera}
\end{equation}
where $\tau$ is a time delay representing the latency of neural
reflexes in human balancing tasks and 
\begin{math}
 R(t) = 1 + \nu\xi(t)
\end{math}
is a random feedback gain, $\xi(t)$ is standard Gaussian white noise,
and $\nu$ represents the strength of the noise.  Note that a stick of
length $l$ and constant linear density is modeled as $\alpha=3g/(2l)$,
where $g$ is the gravitational acceleration.

 Assuming that $\theta\approx\Dx,\; |\Dx|\ll 1$, we obtain
the linearized version of \reqn{cabrera}:
\begin{equation}
 \ddDx + \gamma\dDx - \alpha\Dx
  + \beta\,R(t)\Dx(t-\tau) = 0
  \leqn{bormann04}
\end{equation}
which can be interpreted as an equation of motion of the relative
displacement $\Dx:=x_T(t)-x_M(t)$, where $x_T$ is the displacement of
the upper end of the stick and $x_M$ is the displacement of the lower
end of the stick in the balancing task. It has been reported
\cite{Bormann04} that $x_T(t)$ and $x_M(t)$ are governed by the
following equations:
\begin{equation}
 \ddot x_T + \gamma\dot x_T
    = \alpha \Dx(t),\quad 
  \ddot x_M + \gamma\dot x_M
    = \beta\,R(t)\Dx(t-\tau). 
 \leqn{bormann04:TM}
\end{equation}

\begin{figure}[t]
 \centering
 \includegraphics[width=.8\hsize]{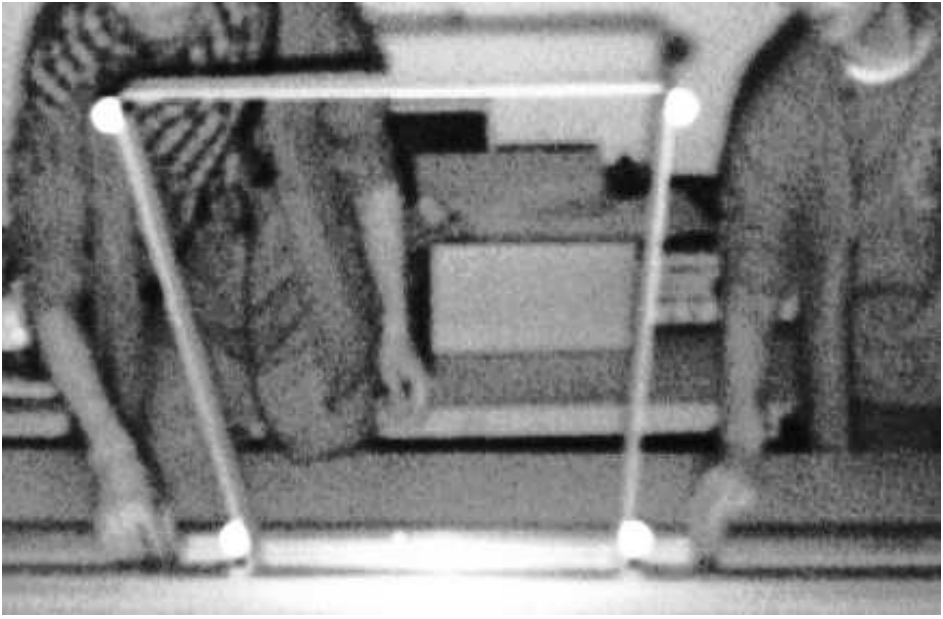}
 \caption{An intuitive example representing a coupled human balancing
 task.}  \lfig{cip_photo}
\end{figure}

\subsection{Coupled balancing tasks}

In the present paper, we investigate the type of stability that can
arise if two subjects that can be balanced independently are linked with
a mechanical structure. \Rfig{cip_photo} shows an intuitive example
representing this situation, in which each subject manipulates one of
the sticks at the lower end along a mechanical slider with the goal of
maintaining the stick in the upright position. Let $q_{Ti}$ and $q_{Mi}$
($i=1,2$) be the horizontal displacement of the upper and lower ends,
respectively, of the $i$th stick.  Then, the presence of the connecting
rod can be described by the distance $l:=q_{T1}-q_{T2}$ maintained by
the fixed length of the rod.  Note that this constant length yields the
following equalities:
\begin{equation}
 \dot q_{T1}=\dot q_{T2}=:\dot q_{T},\quad 
 \ddot q_{T1}=\ddot q_{T2}=:\ddot q_{T}.
 \leqn{dqT}
\end{equation}

For a simpler description of this coupled task, we propose
the following model:
\begin{align}
 &2\ddot q_T + 2\gamma\dot q_T 
    = \alpha \Dq_1(t) +  \alpha \Dq_2(t), \notag
 \\
 &\ddot q_{Mi} + \gamma\dot q_{Mi} = u_i(t,\tau),\notag
 \\
 &u_i(t,\tau) := \beta\,R_i(t)\Dq_i(t-\tau)
 \quad(i=1,2)
\leqn{sss08:TM}%
\end{align}%
where $\Dq_i := q_T-q_{Mi}$ ($i=1,2$) represents the horizontal
displacement between the upper and lower ends, respectively, of the
$i$th stick, as obtained in independent balancing, and $R_i(t) = 1 +
\nu\xi_i(t)$ ($i=1,2$) are independent random feedback gains, where
$\xi_i$ ($i=1,2$) are mutually independent standard Gaussian white
noises. The second equation of \reqn{sss08:TM} can be reduced to the
following relative form:
\begin{equation}
  \ddDq_i + \gamma\dDq_i
    - \frac12 \alpha \Big( \Dq_1 + \Dq_2 \Big)
      +u_i(t,\tau) = 0 \quad (i=1,2).
 \leqn{sss08:TMc}
\end{equation}

\begin{figure}[t]
 \centering\par
 \includegraphics[width=\hsize]{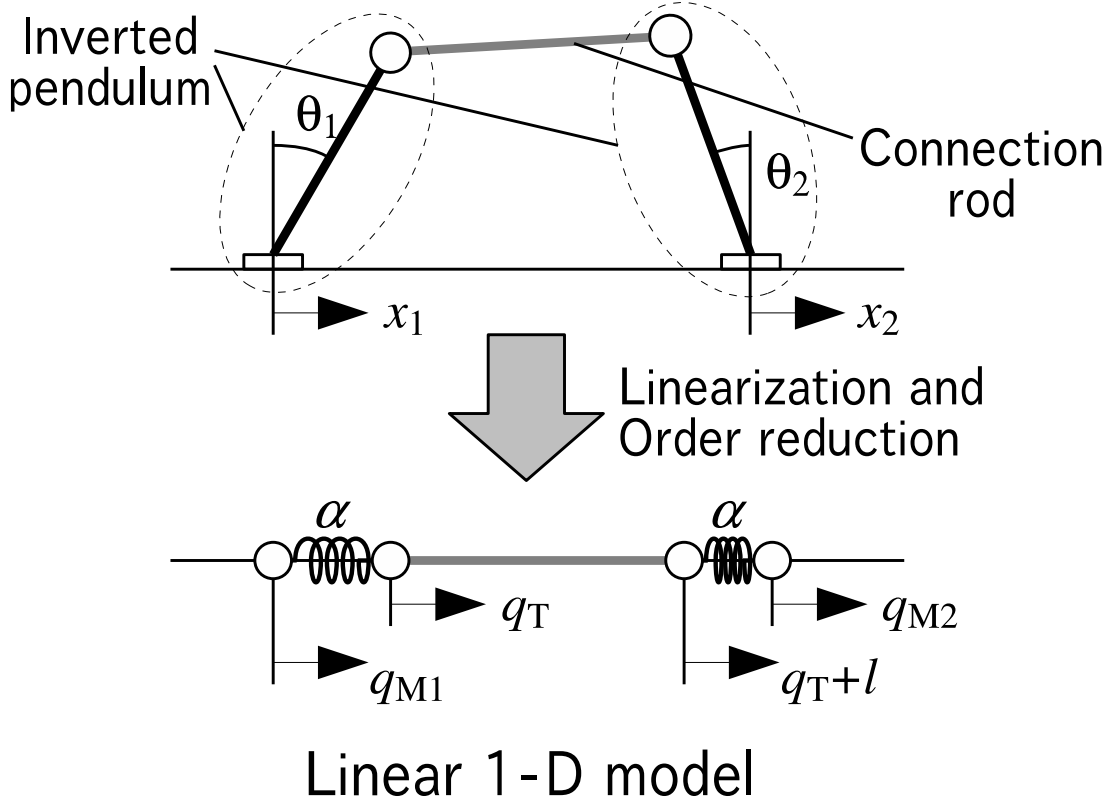}
 \caption{Order reduction and linearization of the CIP model.}
 \lfig{1dmodel}
\end{figure}

The proposed model \reqn{sss08:TM} provides a linearized reduced-order
counterpart of the CIP model \cite{JSDD08}, as shown schematically in
\rfig{1dmodel}. In other words, by assuming $|\theta_i|\ll 1$, the
gravitational restoring force proportional to $\sin\theta_i$ in the CIP
model is approximated as a linear spring force with a negative
coefficient $\alpha<0$.  This setup is expected to provide the simplest
model of coupled human balancing tasks.  Note that the coupled model
\reqn{sss08:TMc} coincides with the single model \reqn{bormann04} if
$\Dq_1=\Dq_2$ and $\dDq_1=\dDq_2$.  In the following, we choose
$\gamma=50$, $\alpha=22$, $\nu=0.6$, and $\tau=0.1$.  For convenience,
we refer to $\Dx$ and $\Dq_i$ as {\em balancing errors}.

 The linearized model \reqn{sss08:TM} conserves the essential
 instability of the stick in the upright position originally described
 in \reqn{cabrera}.  Although this linear approximation becomes invalid
 as the balancing errors $\Dx$ and $\Dq_i$ increase, these errors remain
 sufficiently small during the balancing task of maintaining the stick
 near the upright position.

\section{Effects of Coupling in the Balancing Model}

\subsection{On-off intermittency}

\begin{figure}[t]
 \centering
 \includegraphics[width=\hsize]{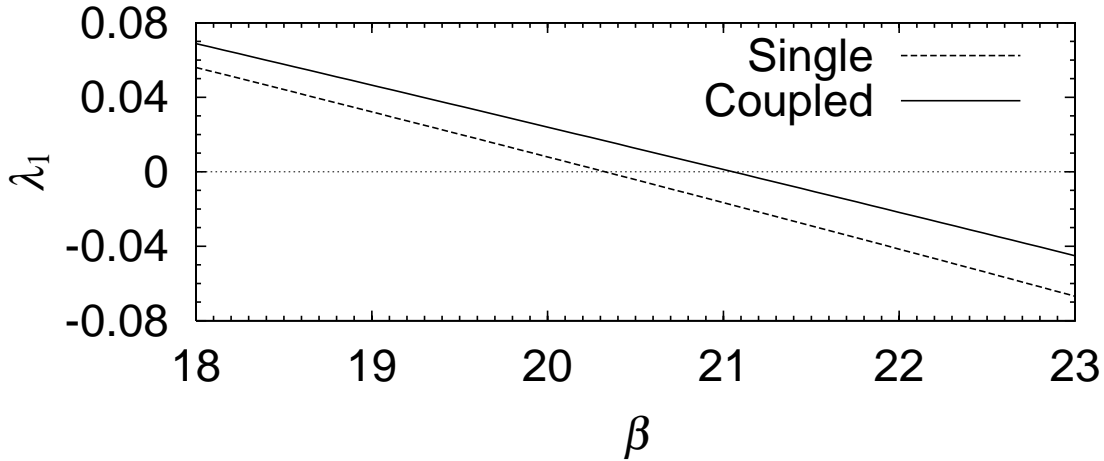}
 \caption{Largest Lyapunov exponent $\lambda_1$ for the single and
 coupled balancing tasks as a function of the feedback gain $\beta$.}
 \lfig{lLe}
\end{figure}

The human stick balancing task exhibits on-off intermittency with
respect to the balancing error \cite{Cabrera02,Bormann04}, and the
on-off intermittency occurs when the largest Lyapunov exponent
$\lambda_1$ is slightly larger than zero \cite{Shankar96}.
 Letting $\bi q=(q_1,\dot q_1,q_2,\dot q_2)$ be the solution of
 \reqn{sss08:TM} and letting $\Delta\bi q$ be the infinitesimal distance
 from the equilibrium point, we obtain the largest Lyapunov exponent
 $\lambda_1$ defined by \cite{Shankar96},
 \begin{equation}
  \lambda_1 = \lim_{t\to\infty}\frac1t\log\|\Delta\bi q(t)\|
 \end{equation}
 where $\|\,\Delta\bi q\|$ represents a Euclid norm of the vector
 $\Delta\bi q$.

\Rfig{lLe} shows the largest Lyapunov exponent $\lambda_1$ of $\Dx$ and
$\Dq_1$ as numerically calculated from \reqn{bormann04} and
\reqn{sss08:TM}, respectively. The results for $\Dq_2$ are omitted due
to their similarity to the results for $\Dq_1$.  We choose
$\lambda_1=5\times10^{-5}$ to be slightly larger than zero in order to
produce on-off intermittency of the balancing errors $\Dx$ in
\reqn{bormann04} and $\Dq_1$ in \reqn{sss08:TM}, as shown in \rfig{Be}.

\begin{figure}[t]
 \centering
 \includegraphics[width=\hsize]{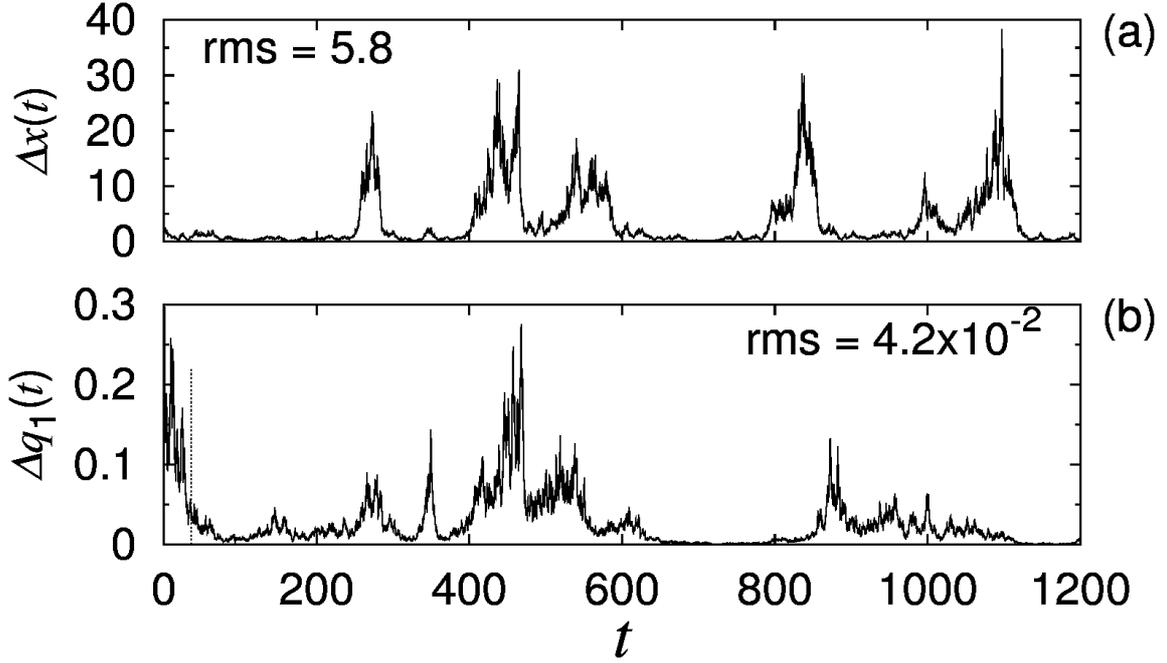}
 \caption{Balancing error $\Delta x(t)$ of the single system
 \reqn{bormann04} for $\lambda_1=5\times10^{-4}$ ($\beta=20.306$)
 and $\Delta q_1(t)$ of the coupled system \reqn{sss08:TM} for
 $\lambda_1=5\times10^{-4}$ ($\beta=21.032$).}
 \lfig{Be}
\end{figure}

\begin{figure}[t]
 \centering
 \includegraphics[width=\hsize]{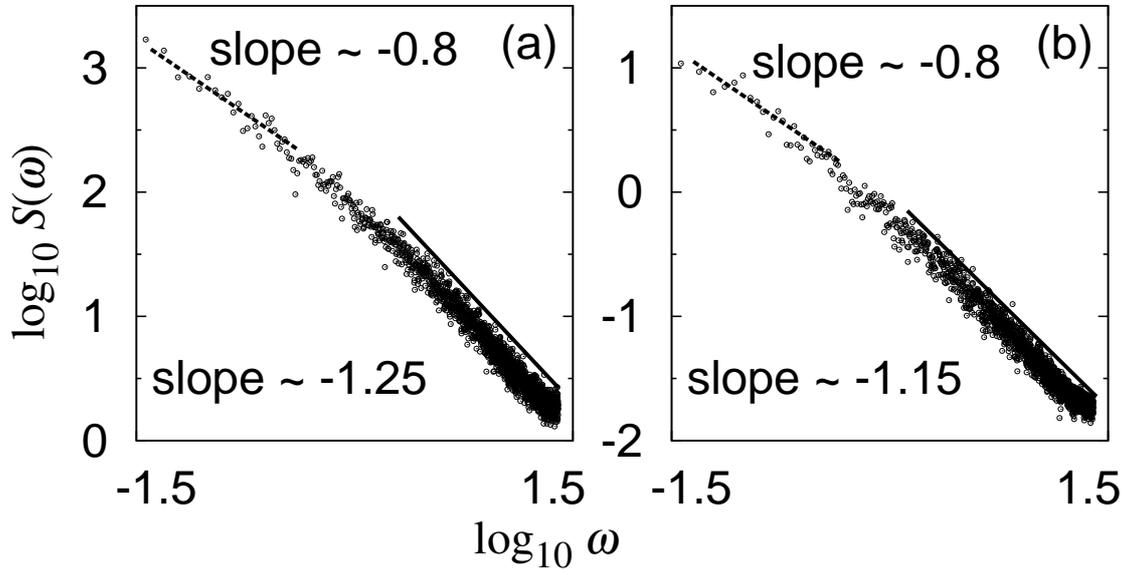}
 \caption{Scaling laws in the power spectra of (a) the single balancing
 task in \reqn{bormann04} for $\lambda_1=5\times10^{-4}$
 ($\beta=20.306$) and (b) the coupled balancing task in \reqn{sss08:TM}
 for $\lambda_1=5\times10^{-4}$ ($\beta=21.032$).}
 \lfig{pwsp}
\end{figure}

Figures \ref{fig:pwsp}(a) and \ref{fig:pwsp}(b) show double logarithmic
plots of the power spectra of the balancing errors $\Dx$ and $\Dq_1$,
respectively. The results for $\Dq_2$ are omitted due to their
similarity to the results for $\Dq_1$.  Both results have two linear
slopes representing two power law regimes, which is in good agreement
with previous balancing experiments
\cite{Cabrera02,Cabrera04,Bormann04}, in which the two slopes were
interpreted as a sign of on-off intermittency \cite{Shankar96}.  Based
on these results, it can be concluded that the coupling term does not
change the scaling law typical of the on-off intermittency of balancing
tasks.

\subsection{Coupling-induced stability}

\begin{figure}[t]
 \centering\par
 \includegraphics[width=\hsize]{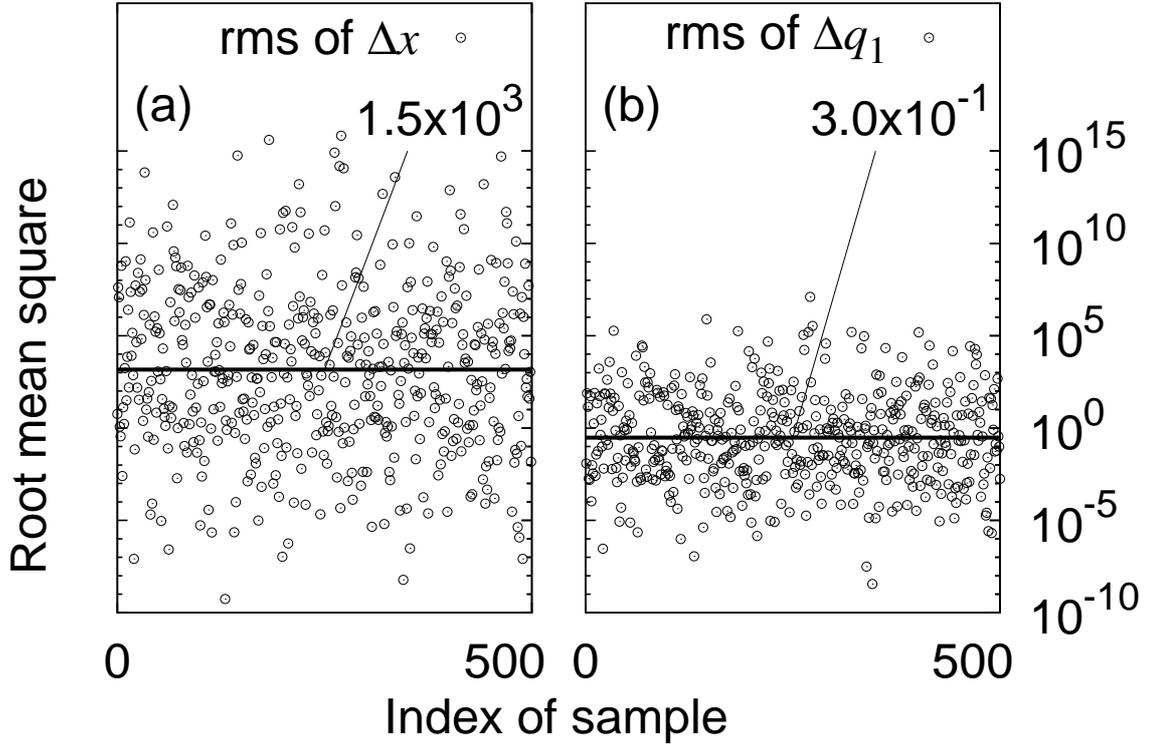}
 \caption{Logarithmic plot of the root mean squares of balancing error
 generated by different samples of the white noise $\xi_i(t)$. Solid
 lines represent the respective mean values of the logarithmic plots.}
 \lfig{rms}
\end{figure}

The most remarkable effect of the coupling in \reqn{sss08:TM} is that
the balancing error is drastically suppressed by the coupling, as shown
in Figs. 4(a) and 4(b). In other words, the error $\Dq_1$ for coupled
balancing is as low as 1\% of $\Dx$ of that of single balancing in terms
of both maximal values and root mean square (RMS) values.  \Rfig{rms}
shows logarithmic plots of 500 realizations of the RMS values. The mean
RMS values, represented by the solid lines, are $1.5\times 10^{3}$ for
single balancing and $3.0\times 10^{-1}$ for coupled balancing.  The
results for $\Dq_2$ are omitted due to their similarity to those for
$\Dq_1$.  \Rfig{rms} shows that the coupling term reduces the balancing
error $\Dq_1(t)$ to 0.02\% of $\Dx$ for single balancing.  This result
implies that the mechanical coupling structure can improve the stability
of the amplitude in human balancing tasks.

In order to examine the velocities of the balancing errors, the ratio of
the probability density $p(\dDq_1)$ divided by $p(\dDx)$ is shown in
\rfig{velo}, where $\dDq_1$ and $\dDx$ are the velocity errors for the
coupled balancing and the single balancing tasks,
respectively. \Rfig{velo} shows that $p(\dDq_1)$ is approximately twice
as high as $p(\dDx)$ near the origin. Since, in the linear
approximation, the velocity errors $\dDx$ and $\dDq_i$ are proportional
to the respective slant angles of the sticks, the results suggest that
the probability of maintaining constant angles is nearly doubled by the
mechanical coupling structure.

\begin{figure}[t]
 \centering\par
 \includegraphics[width=\hsize]{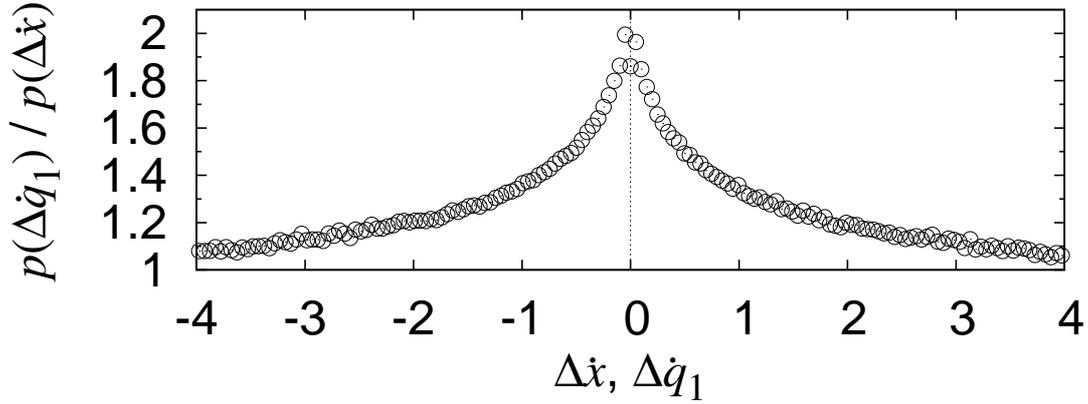}
 \caption{Ratio of the respective probability densities of the
 velocities $\dDq_1$ and $\dDx$ for coupled and single
 balancing.}  \lfig{velo}
\end{figure}

Based on these results, the mechanical coupling structure may reduce the
balancing error in terms of both amplitude and velocity under suitable
conditions.
 Although it is beyond the scope of the present study, one possible
 explanation of this coupling-induced stability might be that the
 coupling causes some stochastic averaging effect, whereby the two
 independent fluctuations $\xi_i(t)$ $(i=1,2)$ cancel each other through
 the coupling constraint. This canceling process, however, occurs in a
 stochastic manner so that further stochastic analysis related to the
 noise-induced order \cite{NIO} in the field of statistical physics will
 be helpful in explaining the coupling-induced stability.

\subsection{Coupling-induced sensitivity}
\lsec{CIS}

It has been reported that near stability boundaries, parametric noise
can allow time-delayed feedback controllers to produce corrective
movements on a time scale shorter than that of the delayed feedback
\cite{Cabrera02}. In order to observe the effect of this improvement in
sensitivity as a function of time, we consider the short-time
cross-correlation coefficient (STCC) in the following manner.  Letting
$x(t), y(t)$ be a pair of time series to be compared and letting $\Dt$
be the length of the time interval of short-time averaging, we define
the STCC as follows:
\begin{align}
 &R(x,y;\tau)(t) = \frac{C(x-m_x,y-m_y;\tau,t)}{\sigma_x\sigma_y},
 \leqn{STCC}\\
 \intertext{where}
 &C(x,y;\tau,t) := \ave{x(s)y(s+\tau)}_{[t,t+\Dt]},\notag\\
 &m_x:= \ave{x(s)}_{[t,t+\Dt]},\;
 \sigma_x:= \ave{(x(s)-m_x)^2}_{[t,t+\Dt]}^{1/2}\notag
\end{align}
where $\ave{X(s)}_{[a,b]}:=(b-a)^{-1}\int_a^b X(s)ds$ is the temporal
average of $X(s)$ over the time interval $[a,b]$.  

In the following, we draw a comparison between the velocities of the
upper end and the lower end of each stick by using the STCC in
\reqn{STCC}.  More specifically, we focus on $R(\dot x_T,\dot x_M)$ for
the single system \reqn{bormann04:TM} and $R(\dot q_T,$ $\dot q_{M1})$
($i=1,2$) for the coupled system \reqn{sss08:TM}.  Moreover, in order to
evaluate the sensitivity of the corrective movements of the feedback
controllers, we also define the first dominant peak points
$\hat\tau_{\dot x}$, $\hat\tau_{\dot q_1}$, and $\hat\tau_{\dot q_2}$ in
$R(\dot x_T,\dot x_M)$ and $R(\dot q_T,\dot q_{Mi})$ ($i=1,2$).  These
peak points are assumed to evaluate the {\it tracking abilities} of the
feedback controllers.

\begin{figure}[t]
\centering
 \includegraphics[width=\hsize]{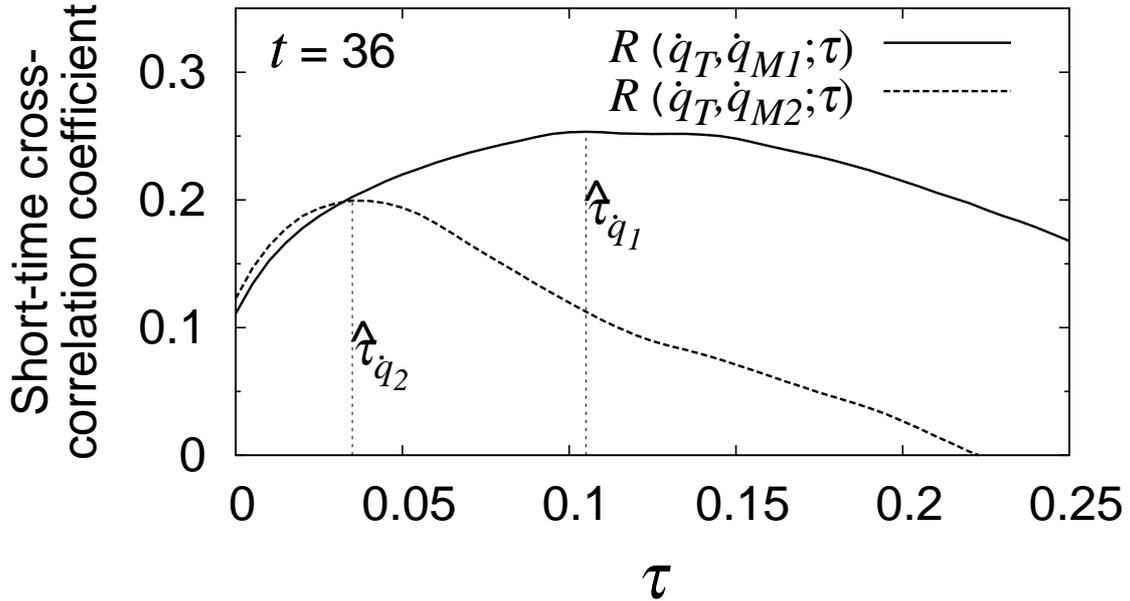}
 \caption{Short-time cross-correlation coefficient at $t=36$.}
 \lfig{scc}
\end{figure}

\Rfig{scc} shows the realization of the STCCs, $R(\dot q_T,$ $\dot
q_{M1})$ and $R(\dot q_T,\dot q_{M2})$ of the coupled system
\reqn{sss08:TM} at $t=36$ for $\Dt = 5$.  In this plot, there are single
peaks at $\hat\tau_{\dot q_1}=0.105$ and $\hat\tau_{\dot q_2}=0.035$.
Since the delay time of the controller is $\tau=0.1$ in our
calculations, on average, the first controller $u_1(t,\tau)$ corrects
the stick movement more slowly than the delay time, whereas the second
controller $u_2(t,\tau)$ performs the corrections in a shorter time than
the delay time.
Regarding $\hat\tau_{\dot q_1}$, $\hat\tau_{\dot q_2}$ as the tracking
ability of the controller, the second controller can
track the stick movement $0.105/0.035=3$ times faster than the first
controller, which indicates that symmetrically placed controllers $u_1$
and $u_2$ having the same specifications can develop asymmetric tracking
abilities over short time scales.

\begin{figure}[p]
 \centering
 \includegraphics[width=\hsize]{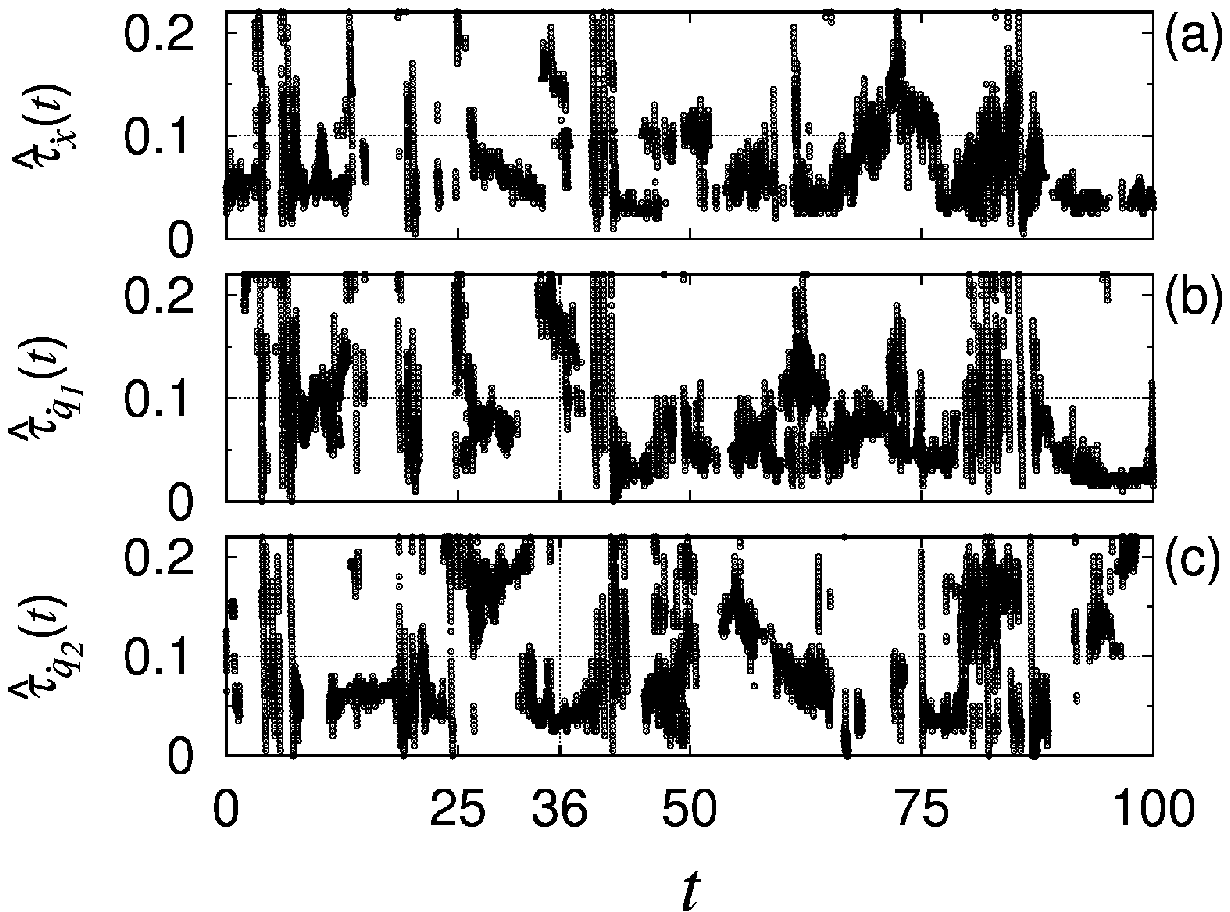}
 \caption{First peak points
 $\hat\tau_{\dot x}(t)$,
 $\hat\tau_{\dot q_1}(t)$,
 and $\hat\tau_{\dot q_2}(t)$ 
 of the short-time cross-correlation coefficients
 $\STCC{\dot x_T}{\dot x_M}$, 
 $\STCC{\dot q_{T}}{\dot q_{M1}}$, and 
 $\STCC{\dot q_{T}}{\dot q_{M2}}$, respectively, at time $t$.}
 \lfig{sccpeak}
 \centering\par\vskip\floatsep
 \includegraphics[width=\hsize]{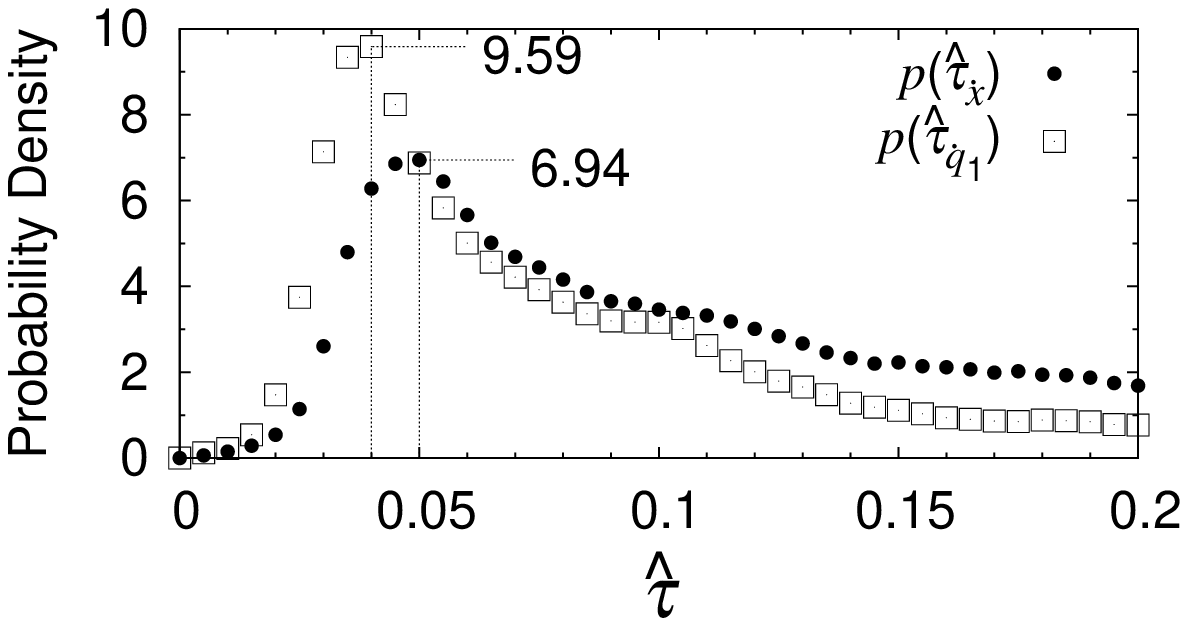}
 \caption{Probability densities of the first peak point 
 $\hat\tau_{\dot x}(t)$ for the single case 
 and $\hat\tau_{\dot q_1}(t)$ for the coupled case as
 constructed from 100 samples in the time interval $t\in[0:1200]$.}
 \lfig{sccpeak:pdf}
\end{figure}

The peak points $\hat\tau_{\dot x}$ and $\hat\tau_{\dot q_i}$ ($i=1,2$)
are plotted in \rfig{sccpeak} as a function of time $t$.  The plots are
numerically constructed from single realizations of numerical solutions
of \reqn{bormann04:TM} and \reqn{sss08:TM}.  The peak points randomly
fluctuate over time, forming intermittent clusters of points.
The sensitive region in which the peak point is smaller than the delay
time is confirmed to be of sufficient length for physical observations.
However, the effects of coupling are not clearly observable in
\rfig{sccpeak}. Therefore, the differences between the results (a), (b),
and (c) in \rfig{sccpeak} in these plots are slight.

In order to evaluate the effects of coupling on tracking ability, the
probability densities of the peak points $\hat\tau_{\dot x}$ and
$\hat\tau_{\dot q_1}$ are shown in \rfig{sccpeak:pdf} as averaged over
100 realizations over the time interval $[0,1200]$ of the numerical
solutions of \reqn{bormann04:TM} and \reqn{sss08:TM}. The results for
$\hat\tau(\dot q_2)$ are omitted because it is quite similar to that for
$\hat\tau(\dot q_1)$. It is clearly observed in \rfig{sccpeak:pdf} that
the density of coupled balancing, represented as small rectangles,
produces a simple peak, which is 38\% higher than that for single
balancing, where the peak is placed at a time scale of $\hat\tau$, which
is 20\% shorter than that for single balancing.

Based on these results, it can be concluded that the mechanical coupling
structure increases the probability of the occurrence of faster
corrective movements and improves the tracking ability of the
controllers.
 One possible interpretation could be that this noise-induced
 sensitivity is a version of the stochastic resonance\cite{SR} (SR), a
 random phenomenon modeled by a heavily damped particle moving randomly
 in a double well potential. The SR approach has been applied to single
 human balancing\cite{SR:balance}.  However, the SR approach to the
 coupled human balancing remains unclear and will be discussed later
 herein.

\section{Experiment of Coupled Visuomotor Tracking on a Computer Screen}

\subsection{Experimental setup}

We perform an experiment in which the numerical pseudo-neural
controllers in \reqn{sss08:TM} are replaced by natural human balancing
tasks, as shown in \rfig{setup}(a).  In practice, the variable $x_{M}$
in the single model \reqn{bormann04:TM} is replaced with the movement of
a mouse manipulated by a subject, and $q_{M1}$ and $q_{M2}$ in
\reqn{sss08:TM} are replaced by those of two subjects.  Each subject is
presented with combinations of thick and thin lines on a computer
screen, as shown in \rfig{setup}(b). The thick lines represent the upper
ends $x_T$ and $q_{Ti}$ as calculated from the numerical models
\reqn{bormann04:TM} and \reqn{sss08:TM}, while the thin lines represent
the lower ends $x_M$ and $q_{Mi}$ manipulated by the subjects.  The
screen resolution is $1,200\times 600$ (pixels), where the range of the
displacement $[-3,3]$ in the numerical model maps to the horizontal
range of pixels $[1,1,200]$ on the screen.  The movements performed by
the subjects $x_M$ and $q_{Mi}$ are recorded and substituted into the
numerical models \reqn{bormann04:TM} and \reqn{sss08:TM} with a sampling
rate of 50 Hz, and the set of lines on the screen is animated at the
same rate.

\begin{figure}[p]
  \centering\par
  \includegraphics[width=.7\hsize]{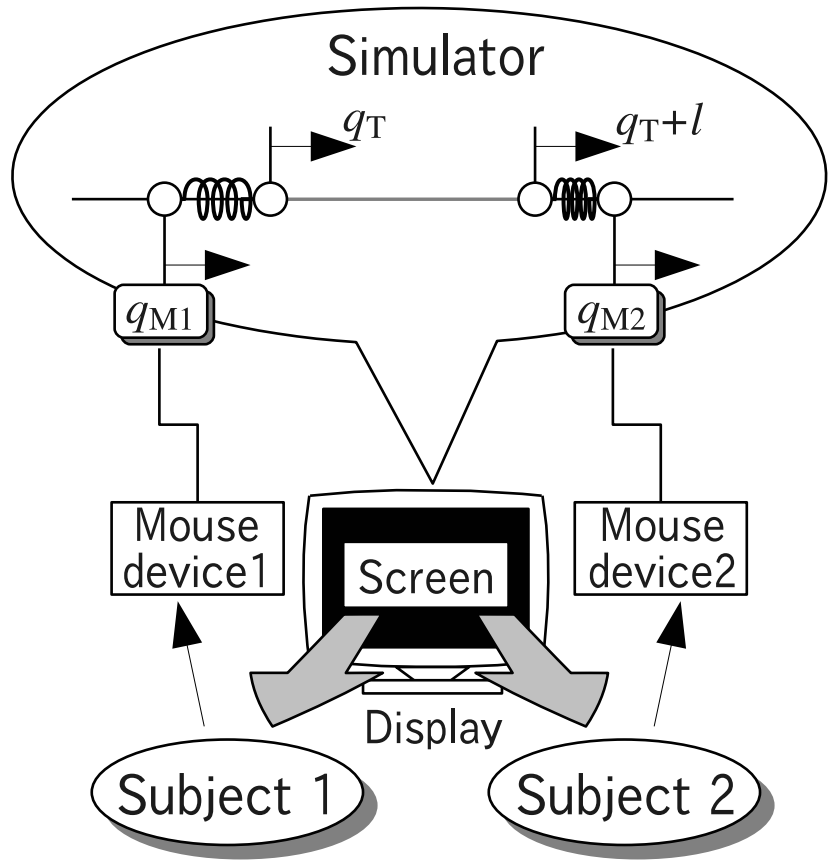}
 \\[\floatsep]{(a)~~Experimental device.}
  \centering\par\vskip\floatsep
  \includegraphics[width=.85\hsize]{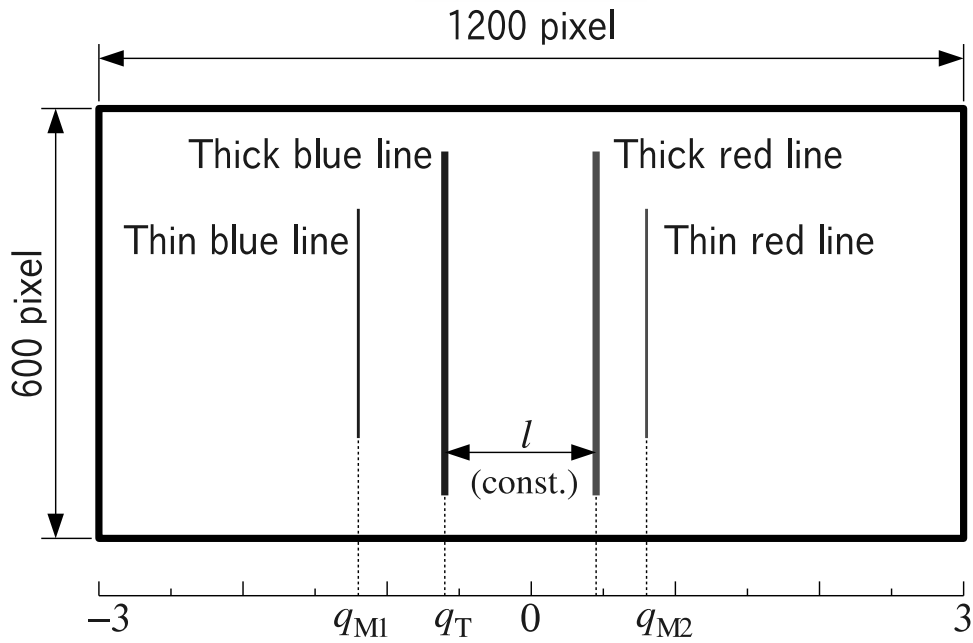}
 \\[\floatsep]{(b)~~Screen design.}
  \caption{Experimental setup of the coupled human balancing tasks.}
  \lfig{setup}
\end{figure}

The experimenter issues the following instructions to the subjects:
\begin{itemize}
 \item Each subject should collaborate with his/her partner in order to
       maintain the assigned pendulum, provided as an overhead view on
       the computer screen, in the upright position by manipulating the
       lower end of the pendulum, which is represented as a thin line.
 \item The upper ends, represented as thick lines, are assumed to be
       connected by a rigid rod in such a way that a constant distance
       is maintained between the thick lines.
 \item The experiment begins after a countdown performed by the
       experimenter, and ends after ten minutes or when any of the lines
       leaves the visible range.
 \item The subject can abort the experiment at anytime.
\end{itemize}
According to these instructions, after the countdown, the numerical
simulation is started and the experiment begins.  The experiment ends
after ten minutes or when any of the lines $x_T$, $q_{Ti}$, and $x_M$,
$q_{Mi}$ leaves the visible range $[-3,3]$.  The initial positions of
the animated lines, i.e., the initial values of the models, are set to
$x_{T}(0)=-0.5$, $q_{M1}(0)=-0.6$, $q_{M2}(0)=0.6$, $\dot x_{T}(0)=\dot
x_{M}(0)=0$, $q_{T}(0)=-0.5$, $q_{M1}(0)=-0.6$, $q_{M2}(0)=0.6$, and
$\dot q_{T}(0)=\dot q_{M1}(0)=\dot q_{M2}(0)=0$.  The model parameters
are chosen to be $\alpha=\beta=21$, $\gamma =50$, and $l=1$.

The six subjects were healthy volunteers aged 21 to 24 years. The
experiments were performed according to the principles of the
Declaration of Helsinki and informed consent was obtained. The
experimental protocol was approved by the Bioethics Committee of
Utsunomiya University. The subjects were labeled A, B, C, D, E, and F,
and three pairs (A,B), (C,D), and (E,F) are considered in the coupling
task.

\begin{figure}[t]
  \centering\par
  \includegraphics[width=\hsize]{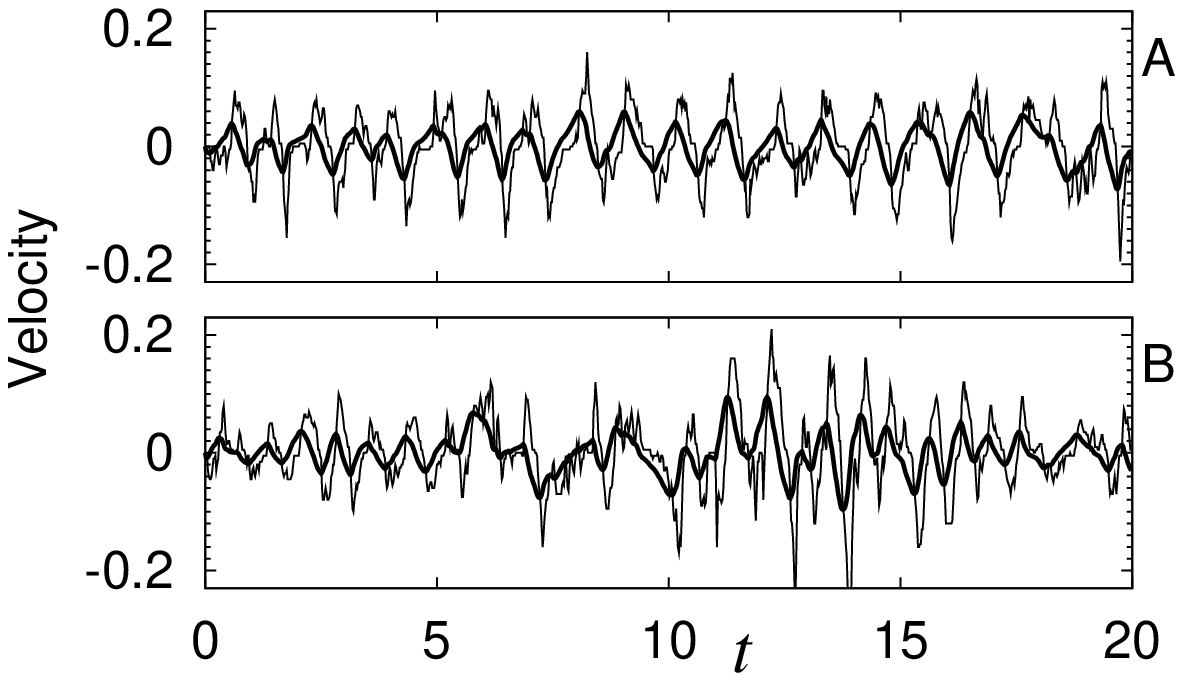}\\
  \includegraphics[width=.9\hsize]{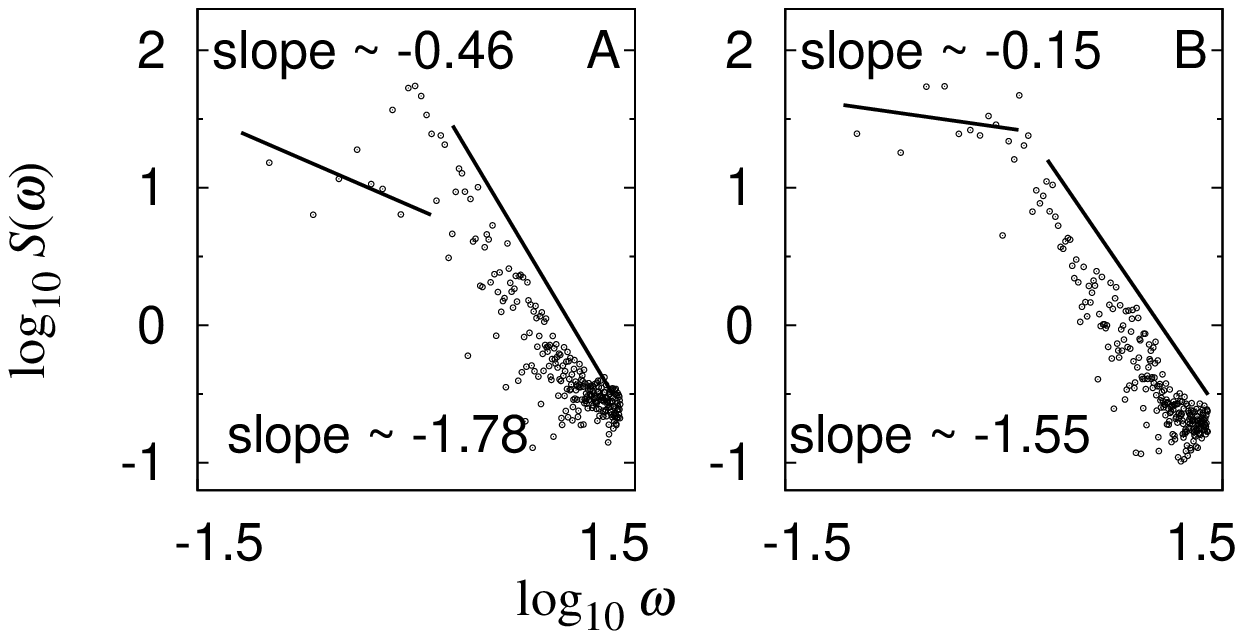}
  \caption{Experimental velocities of the target $\dot x_T$ (thick line)
  and of the subject's movement $\dot x_M$ (thin line), and power
  spectra of the balancing error $\Delta x$ for single balancing
  performed independently by subjects A and B.}
  \lfig{move_s}
\end{figure}

\subsection{Experimental results}

First, we focus on subjects A and B and the pair (A,B).  \Rfig{move_s}
shows the experimental results for single balancing performed
independently by subjects A and B.  The thick line, which corresponds to
the velocity $\dot x_T$ of the target (the upper end of the stick),
slightly precedes the thin line, which corresponds to the subject's
velocity $\dot x_M$, due to a combination of time delays from neural
reflexes and computer processing.  Moreover, for the case in which
balancing is performed by both subjects, the power spectrum $S(\omega)$
of the balancing error $\Dx$ exhibits a power law behavior typical of
neural controllers, which have two different power laws with an exponent
of $\approx -1/2$ in the lower-frequency range \cite{Cabrera02}.

\begin{figure}[t]
  \centering\par
  \includegraphics[width=\hsize]{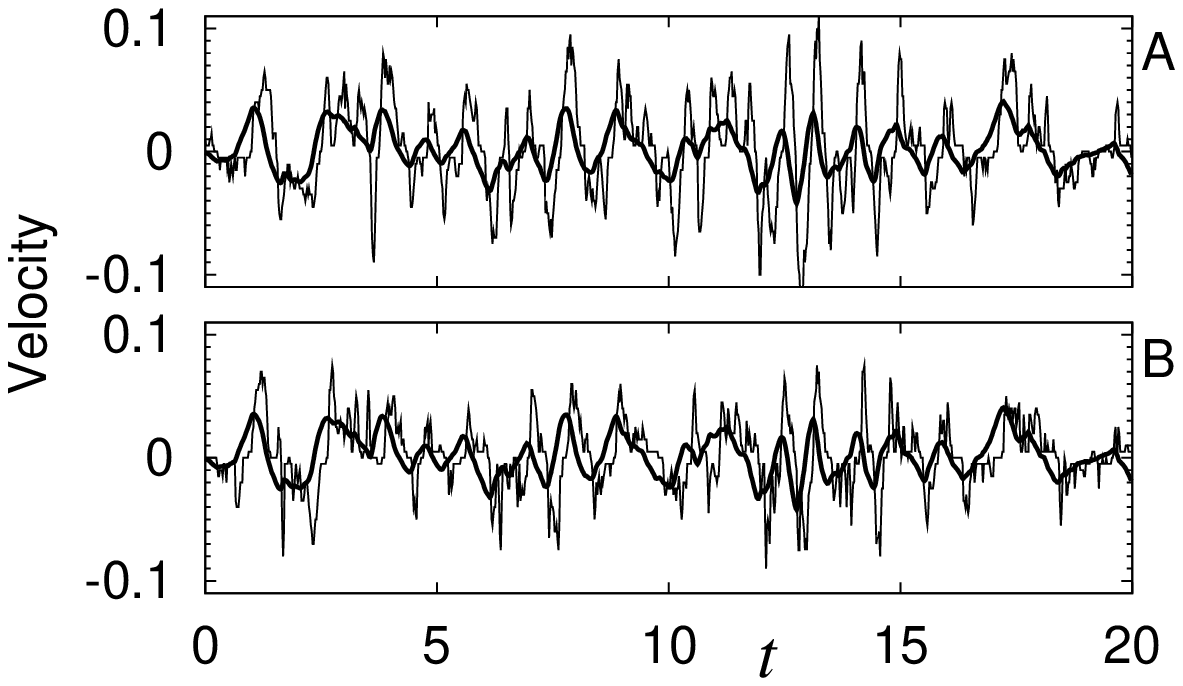}\\
  \includegraphics[width=.9\hsize]{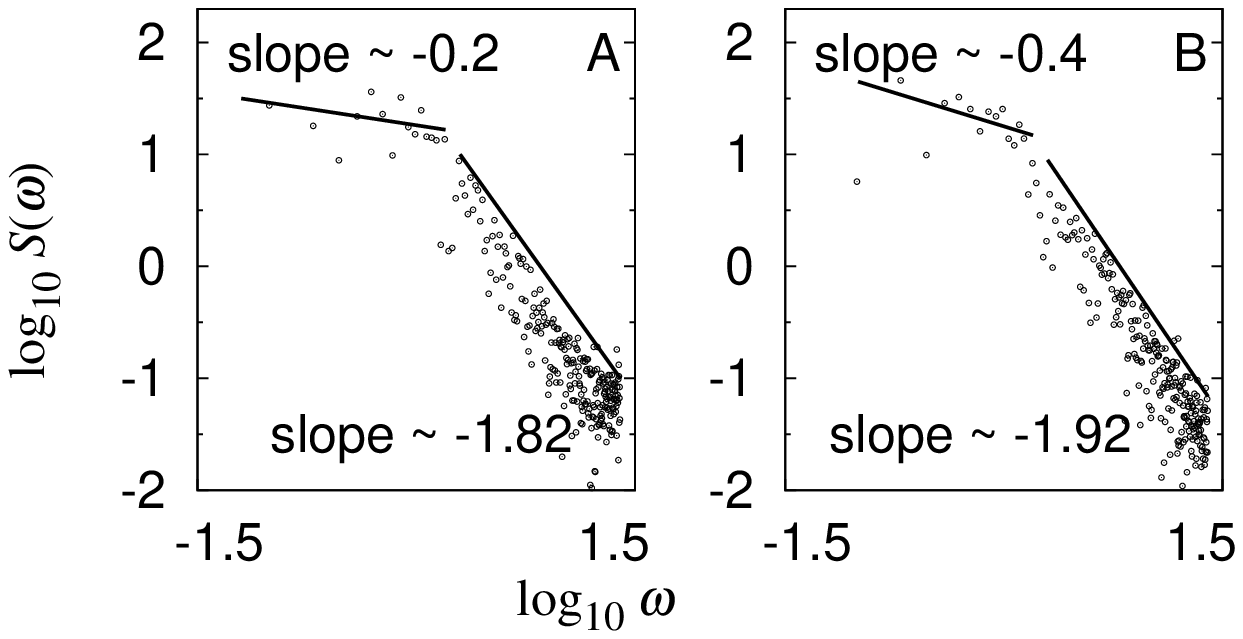}
  \caption{Experimental velocities of the target $\dot q_{T}$ (thick
  line) and of the subject's movement $\dot q_{Mi}$ (thin line), and
  power spectra of the balancing error $\Dq_i$ for coupled
  balancing by the pair of subjects A ($i=1$) and B ($i=2$).}
 \lfig{move_c}
\end{figure}

On the other hand, \rfig{move_c} shows the experimental results for the
case in which the same subjects are coupled by a rigid rod of length $l$
in the numerical model. Note that, as a result of the equalities
\reqn{dqT}, the velocities of both targets (thick lines) coincide.  The
thick line, which corresponds to the target velocity $\dot q_{T}$,
slightly precedes the thin line, which corresponds to the velocities
$\dot q_{M1}$ and $\dot q_{M2}$ of the subject, and the balancing errors
$\Dq_1$ and $\Dq_2$ are governed by a power law having an exponent of
$\approx -1/2$ in the lower-frequency range.  Based on these results,
the mechanical coupling structure between the two balancing tasks
maintains the time-delay and scaling-law properties typical of the
independently performed single balancing tasks.

\begin{figure}[t]
 \centering\par
 \includegraphics[width=\hsize]{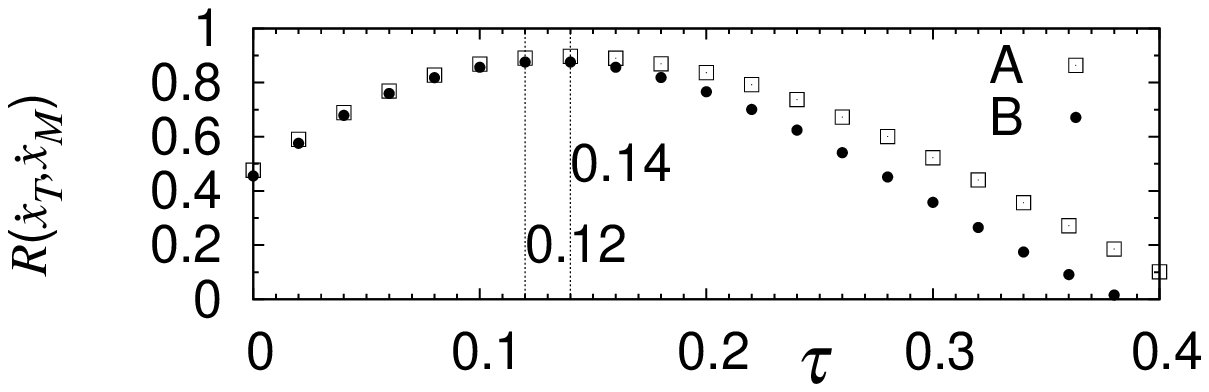}\\
 \caption{Experimental short-time cross-correlation coefficients $R$
 for single balancing performed independently by 
 subjects A and B.}
 \lfig{stcc_s}
 \centering\par\vskip\floatsep
 \includegraphics[width=\hsize]{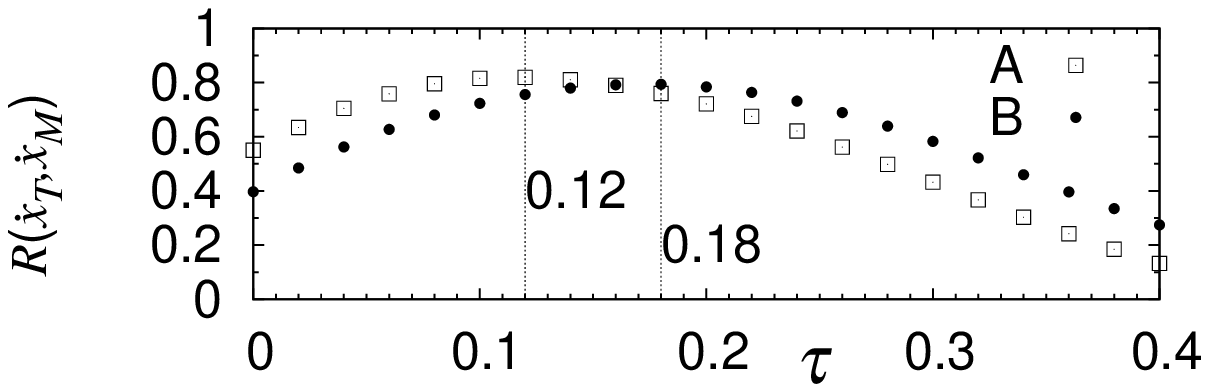}\\
 \caption{Experimental short-time cross-correlation coefficients $R$
 for coupled balancing performed by pair (A,B).}
 \lfig{stcc_c}
\end{figure}

\subsection{Effects of coupling on stability and sensitivity}

As discussed in Section \rsec{CIS}, the correlation time (the first
dominant peak in STCC) allows us to evaluate the sensitivity of the
corrective movements of the subjects, i.e., the tracking ability of the
subjects.  The experimental correlation times for single balancing
performed independently by the two subjects (denoted as A and B) are
shown in \rfig{stcc_s} as peak points $\hat\tau_{\dot x_T}=0.14$, $0.12$
of STCC.  The differences between the correlation times become slightly
larger when the mechanical coupling structure is placed between the two
balancing tasks, as shown in \rfig{stcc_c}, i.e., $\hat\tau_{\dot
x_T}=0.12$, $0.18$.
The correlation times obtained from different trials of the experiment
and their averages are listed in \Rtab{STCC}.  The averages in
\Rtab{STCC} show that the coupling structure increases the individual
differences between the subjects.
In other words, the single balancing tasks yield similar tracking
abilities ($\ave{\hat\tau}=0.132$, $0.136$), whereas the tracking
abilities become asymmetric ($\ave{\hat\tau}=0.128$, $0.168$) when
coupled by the mechanical structure.
Similarly, \rtab{RMS} lists the RMS of the balancing errors and their
averages for estimating the stabilities of the balancing.  The averages
in \rtab{RMS} indicate that the coupling has the opposite effect on the
RMS, as compared with the STCC, i.e., the coupling structure decreases
the individual differences in the RMS of the balancing errors from
$\ave{\text{RMS}}=3.62$ and $5.34$ for the single task to
$\ave{\text{RMS}}=3.02$ and $2.50$ for the coupled task.

\begin{table}[t]
 \centering
 \def\indbox#1{\hbox to5ex{\hfil#1\hfil}}
 \caption{Correlation times at different trials for pair (A,B).}
 \begin{tabular}{c|c|c|c|c|c|c}\hline
  Subject& \multicolumn{5}{|c|}{Index of trial} &Average \\
  &\indbox 1&\indbox 2&\indbox 3&\indbox 4&\indbox 5&
  $\ave{\hat\tau}$\\\hline\hline
  Single:&&&&&& \\
  A &0.12&0.14&0.14&0.12&0.14&0.132 \\
  B &0.14&0.12&0.14&0.14&0.14&0.136 \\\hline
  Coupled:&&&&&& \\
  A &0.12&0.12&0.12&0.12&0.16&0.128 \\
  B &0.18&0.16&0.16&0.18&0.16&0.168 \\\hline
 \end{tabular}
 \ltab{STCC}
 \par\vskip\floatsep
 \caption{RMS of the balancing errors at different trials for pair
 (A,B).}
 \begin{tabular}{c|c|c|c|c|c|c}\hline
  Subject& \multicolumn{5}{|c|}{Index of trial} &Average \\
  &\indbox 1&\indbox 2&\indbox 3&\indbox 4&\indbox 5&
  $\ave{\mbox{RMS}}$\\\hline\hline
  Single:&&&&&& \\
  A &2.4&2.9&3.8&5.5&3.5&3.62	\\
  B &4.9&4.9&5.8&5.0&6.1&5.34	\\ \hline
  Coupled:&&&&&& \\
  A &2.9&3.5&2.4&3.9&2.4&3.02	\\
  B &2.4&2.8&2.6&2.7&2.0&2.50	\\ \hline
 \end{tabular}
 \ltab{RMS}
\end{table}

The results in \rtab{STCC} and \ref{tab:RMS} imply that the coupling
structure increases the individual differences between the tracking
abilities (STCC), while decreasing those of the balancing stabilities
(RMS).  This effect on the six subjects can be seen in \rfig{rmstau},
which shows experimental plots of the correlation time $\tau$ versus the
RMS of the balancing errors over all trials of the single balancing
tasks performed by subjects A, B, C, D, E, and F and the coupled
balancing tasks performed by pairs (A,B), (C,D), and (F,D).  The plots
with filled triangles indicate the results for the single tasks, and the
plots with rectangles indicate the results for the coupled tasks.
The oval is centered at the average point of the corresponding plots and
the radii in the horizontal and vertical directions represent the
standard deviations of the correlation time and of the RMS of balancing
errors, respectively.  A comparison of the single tasks for the solid
oval with the coupled tasks for the dashed oval reveals that the
coupling structure decreases the individual differences in correlation
time, while increasing the individual differences in the RMS.  This
result statistically confirms the coupling effect arising in the two
subjects, as shown in \rtab{STCC} and \ref{tab:RMS}.

One possible explanation for these effects of coupling on the individual
differences might be the separation of roles assumed by the subjects in
the case that requires cooperation.  However, similar asymmetrical
properties can arise in the numerical controller $u_i(t,\tau)$ in the
second equation of \reqn{sss08:TM}, which has no ability to cooperate.
Moreover, the numerical controllers provide feedback of the combination
of their own past states and the shared state, i.e., $\Dq_i=q_T-q_{Mi}$,
in the following form:
\begin{equation}
 u_i = u_i(\Dq_i) \quad \Big(=\beta\,R_i(t)\Dq_i(t-\tau)
         \text{~in \reqn{sss08:TM}}\Big).
\end{equation}
In contrast, the human subjects can also receive visual feedback of
their partner's state via the shared computer screen in \rfig{setup},
and, as a result, their controller model may appear to have the
following structure:
\begin{equation}
 u_i = u_i(\Dq_1,\Dq_2)\quad(i=1,2).
\end{equation}
Developing precise descriptions of such a human controller model will
provide the first step toward exploring the coupling effects, such as
the changes in the individual differences.
 For this purpose, it appears that approaches based on learning theory
 and evolutionary methods\cite{Genci2009-1,Genci2009-2} will be helpful.
However, a detailed explanation of these effects is beyond the scope of
the present paper and will be discussed elsewhere.

\begin{figure}[t]
 \centering\par
 \includegraphics[width=\hsize]{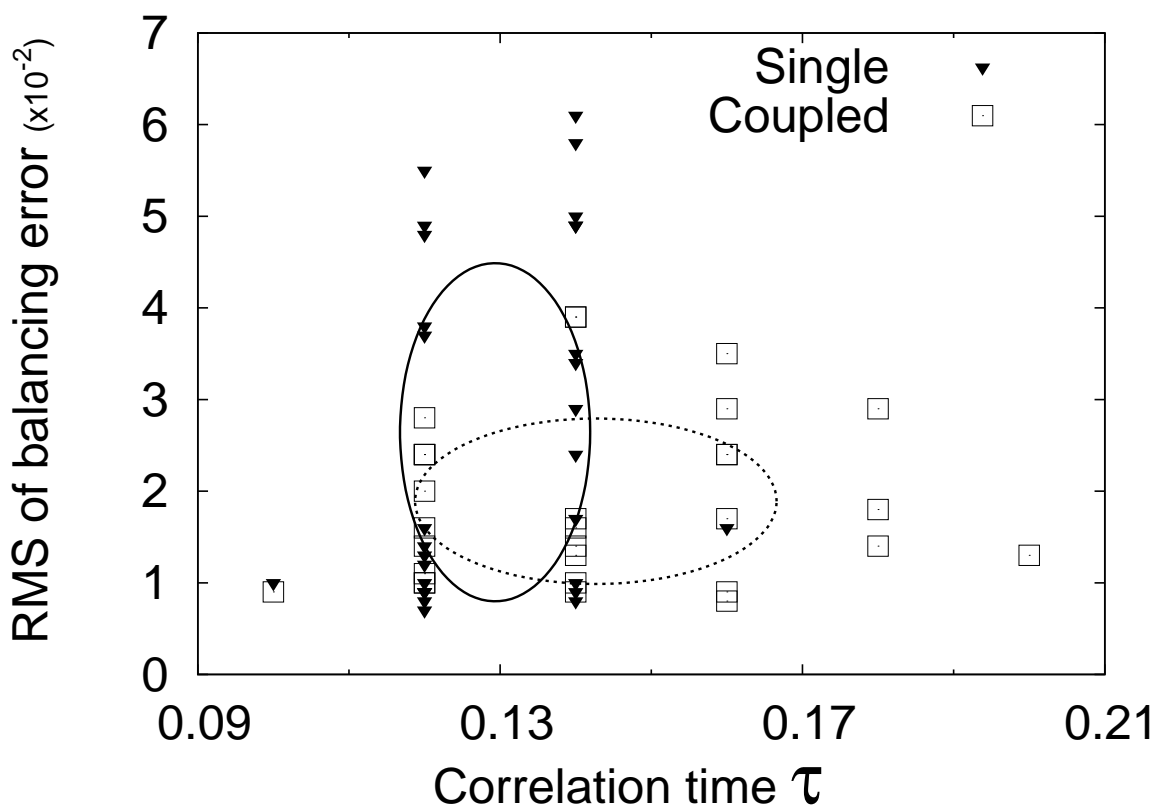}\\
 \caption{Experimental plots of the correlation time $\tau$ versus the RMS
 of the balancing errors over all trials of the single balancing tasks
 performed by subjects A, B, C, D, E, and F and the coupled balancing tasks performed by 
 pairs (A,B), (C,D), and (F,D).}
 \lfig{rmstau}
\end{figure}

\section{Conclusion}

Coupled human balancing tasks are investigated based on pseudo-neural
controllers modeled using time-delayed feedback with random gain.  It is
shown numerically that, compared with the case of single balancing
tasks, the coupling structure increases the stability against balancing
errors in terms of both amplitude and velocity and improves the tracking
ability of the controllers.  We then perform an experiment in which the
pseudo-neural controller in the numerical model is replaced with natural
human balancing tasks carried out using computer mice.
The results reveal that the mechanical coupling structure increases the
individual differences in the tracking abilities between the subjects,
while decreasing the individual differences in the stabilities of the
balancing errors.
The proposed model and experimental method are expected to provide the
simplest means by which to understand the cooperative behavior between
humans sharing mechanical contacts and to provide new insight into
welfare engineering and related fields.  In the future, we intend to
perform model identification of human visuomotor tracking tasks in order
to characterize the trade-off mechanism between tracking abilities and
stabilities produced by coupled human subjects.

\bibliographystyle{ijicic}
\bibliography{CIP}

\end{document}